\documentclass[preprint,showpacs,preprintnumbers,amsmath,amssymb,nofootinbib]
{revtex4}
\usepackage{graphicx}% Include figure files
\usepackage{dcolumn}% Align table columns on decimal point
\usepackage{bm}% bold math
\usepackage{color}
\usepackage{epsfig}
\def\PL #1 #2 #3 {{\it Phys. Lett.} {\bf#1} (#3) #2}
\def\NP #1 #2 #3 {{\it Nucl. Phys.} {\bf#1} (#3) #2}
\def\ZP #1 #2 #3 {{\it Z. Phys.} {\bf#1} (#3) #2}
\def\PRL #1 #2 #3 {{\it Phys. Rev. Lett.} {\bf #1} (#3) #2}
\def\PR #1 #2 #3 {{\it Phys. Rev.} {\bf#1} (#3) #2}
\def\MPL #1 #2 #3 {{\it Mod. Phys. Lett.} {\bf#1} (#3) #2}
\def\RMP #1 #2 #3 {{\it Rev.~Mod. Phys.} {\bf#1} (#3) #2}

\newcommand{\be}{\begin{equation}}
\newcommand{\ee}{\end{equation}}
\newcommand{\ba}{\begin{eqnarray}}
\newcommand{\ea}{\end{eqnarray}}

\newcommand{\e}{\epsilon}

\newcommand{\beqn}{\begin{eqnarray}}
\newcommand{\eeqn}{\end{eqnarray}}
\newcommand{\beqns}{\begin{eqnarray*}}
\newcommand{\eeqns}{\end{eqnarray*}}

\newcommand{\beq}{\begin{equation}}
\newcommand{\eeq}{\end{equation}}
\newcommand{\beqa}{\begin{eqnarray}}
\newcommand{\eeqa}{\end{eqnarray}}
\newcommand{\s}{\slash\hspace{-5pt}}
\newcommand{\nn}{\nonumber \\}

\newcommand{\ep}{\epsilon}
\newcommand{\Res}{{\rm Res}}
\begin{document}
\preprint{FERMILAB-PUB-08-013-T}
\preprint{UH-511-1117-08}
\title{ 
Full one-loop amplitudes from tree amplitudes
}

\author{Walter~T.~Giele}
\email{giele@fnal.gov}
\affiliation{ Fermilab, Batavia, IL 60510, USA }
\author{Zoltan Kunszt}
\email{kunszt@itp.phys.ethz.ch}
\affiliation{Institute for Theoretical Physics, ETH, CH-8093 Z\"urich, Switzerland}
\author{Kirill~Melnikov} 
\email{kirill@phys.hawaii.edu}
\affiliation{Department of Physics and Astronomy,
          University of Hawaii,\\ 2505 Correa Rd. Honolulu, HI 96822}  

\date{\today}
\begin{abstract}
We establish 
an  efficient  polynomial-complexity algorithm for one-loop calculations, 
based on generalized $D$-dimensional unitarity.
It allows 
automated computations  of  both  cut-constructible {\it and}
rational parts of one-loop scattering amplitudes
from on-shell tree  amplitudes.
We illustrate the method by (re)-computing all four-, five- and  six-gluon  
scattering amplitudes in QCD at one-loop.
\end{abstract}
\maketitle

\section{Introduction}

Many complex processes that contain multi-jet 
final states will be observed at
the Large Hadron Collider (LHC).  
Detailed studies of these processes  
will give us  information about the mechanism 
of electroweak symmetry breaking and, hopefully, reveal physics beyond 
the Standard Model.
To assess this information and 
interpret experimental data correctly,  an accurate theoretical description
of processes at the LHC is required.
In principle, Monte-Carlo programs 
based on leading order (LO) computations do provide such a description~\cite{ALPGEN,MADGRAPH,HELAC,CompHEP,AMEGIC}.
However, as experience with the TEVATRON, LEP and HERA data has shown, 
LO predictions often 
give only rough  estimates.
%and do not provide a way to assess theoretical uncertainties.
To extract  maximal information from   data, 
more precise predictions based on  next-to-leading order (NLO)
calculations are required. NLO predictions are also 
instrumental for reliable estimates of theoretical uncertainties 
related to the truncation of the perturbative expansion.

For sufficiently  complicated  final states, computations required 
for the LHC are  
very challenging. In the standard approach one uses perturbative
expansion of scattering amplitudes 
in terms of Feynman diagrams. This gives an  
algorithm suitable for numerical implementation.
However,  even for tree amplitudes,  the number of Feynman
diagrams grows 
faster than factorial with the number of external
particles involved in a scattering process. As a consequence, 
computing time needed to evaluate a scattering amplitude 
at a single  phase-space point, 
grows at least as fast. This means that computational algorithms 
based on expansion in Feynman diagrams are necessarily of 
exponential complexity, an undesirable feature.

In tree-level calculations exponential growth in complexity is avoided by 
employing  recursion relations. These relations
re-use recurring groups of off-shell Feynman graphs in an
optimal manner~\cite{Berends:1987me,Caravaglios:1995cd,Draggiotis:1998gr,BCFrecursion,BCFW}.
The use of recursion relations for tree amplitudes leads
to a computational algorithm of polynomial complexity so that 
computing time grows as some power of  the number of external legs.
Because of that, the problem of evaluating tree amplitudes is 
considered a solved problem by the high-energy 
physics community. As was pointed out in Ref.~\cite{egkLH07}, 
a polynomial-complexity  algorithm 
is not available for one-loop computations; constructing such 
an algorithm is the goal of the present paper. 

When performing NLO computations in the Standard Model,
many difficulties arise. We need 
to calculate  both virtual one-loop corrections
and real emission processes with one additional
particle in the final state.  
Currently, the bottleneck in  NLO calculations for
multi-particle processes  is  the computation of virtual corrections.
The difficulty related to the factorial growth in the number 
of Feynman diagrams is further amplified by a large number of terms 
generated when tensor loop integrals are reduced to scalar integrals.
Nevertheless, thanks to modern computational resources,
standard methods based on Feynman diagrams and tensor integrals reduction~\cite{Passarino:1978jh}   
may be extended brute force to deal with multi-particle processes.  
Striking examples of the success 
of this approach   are recent computations of electroweak corrections 
to $e^+e^- \to~{\rm four~fermions}$ process~\cite{denner} and 
 one-loop
six-gluon scattering amplitudes 
Ref.~\cite{Ellis:2006ss}. Note, however, that 
a single phase space point for six-gluon scattering  is evaluated in 
about nine seconds, which is  10,000
times slower than the evaluation time for four-gluon scattering amplitudes
generated using the same procedure. 
It is clear that further application of brute force approaches 
to yet higher multiplicity processes are becoming  unfeasible.

Unitarity-based methods for multi-loop calculations were suggested as 
an alternative to the expansion in Feynman diagrams long ago. We review 
the status of these calculations in the next section. 
The goal of the present paper is to describe  a polynomial-complexity
computational   algorithm for one-loop amplitudes  
that provides both  cut-constructible
{\it and} rational parts. We start with the idea of generalized 
unitarity in $D$-dimensions and develop it to 
an algorithm amenable to numerical implementation.
Since  one-loop amplitudes are built up from tree amplitudes, albeit 
in higher-dimensional space-time,  the polynomial complexity 
of the algorithm is ensured. The method is flexible and 
can be applied to scattering amplitudes with arbitrary internal 
and external particles. In particular, dealing with massive particles is 
straightforward.

The outline of the paper is as follows. In section II we give 
an overview of the
current status of unitarity techniques.
The structure of 
one-loop scattering amplitudes in
$D$-dimensions is discussed in section III. 
Section IV contains the discussion of the $D$-dimensional residues and  
algebraic extraction of  the coefficients of  master integrals. 
In section V we apply the formalism to gluon scattering amplitudes.  
Numerical results for four-, five- and six-gluon scattering 
amplitudes are reported  in section VI. We conclude in section VII. 

\section{The Status of Unitarity Methods}

Unitarity-based methods for loop calculations were
suggested as an alternative to  the Feynman-diagrammatic expansion long 
ago~\cite{Cutkosky:1960,Diagrammar,vanNeerven:1985xr}. 
It was  argued  that for  gauge theories these methods
lead to higher computational  efficiency than traditional 
methods~\cite{Bern:1994zx,BDKOneloopInt}.
Within unitarity-based methods, computations employ    
tree scattering amplitudes, rather than Feynman diagrams, thereby 
avoiding many complications.
Unitarity cuts factorize  one-loop amplitudes 
into products of  tree amplitudes. Therefore, in numerical implementations
computing time grows with the number of unitarity cuts, rather than 
Feynman diagrams, and depends on  the efficiency 
of algorithms employed for evaluating  tree amplitudes.

The unitarity-based approach gives a complete description of one-loop 
scattering amplitudes  if it is applied in $D$ dimensions.  
Any dimensionally regulated
multi-loop Feynman integral is  fully reconstructible from  
unitarity cuts~\cite{vanNeerven:1985xr}.
Clearly,  in this case we have to associate   
$D$-dimensional momenta and polarization vectors
with  each  cut on-shell line, to obtain one-loop amplitudes~\cite{bernmorgan}.

The first successful application  of unitarity-based techniques for 
one-loop computations employed a 
four-dimensional variant of the unitarity-based 
approach, where four-dimensional states were associated with each 
cut line~\cite{Zqqgg}. 
In analytic calculations,  such a procedure 
has the advantage of allowing  full use of the spinor-helicity formalism.
In this way, however, we obtain only  the so-called cut-constructible part of 
the full one-loop amplitude.  
The missing part is referred to as the rational part and 
has to be determined by other methods. 
In particular, in  supersymmetric theories,
rational parts are known to vanish. In other cases, 
rational parts can be fixed by using factorization properties of 
one-loop  amplitudes in collinear limits~\cite{Zqqgg}.

An important step in developing unitarity-based methods was made in 
Ref.~\cite{BCFgeneralized} where the idea of generalized unitarity 
was introduced.  In particular, it was shown that 
coefficients of four-point scalar 
integrals for multi-gluon processes 
can be calculated  using  quadruple cuts.
This approach  is also suitable for  direct numerical implementation. 
A quadruple cut factorizes an one-loop amplitude into a product of four 
tree amplitudes. From unitarity constraints,  
we derive two complex solutions 
for the loop-momentum. The product of  tree amplitudes can be evaluated 
using these solutions. Coefficients of scalar four-point functions 
are obtained    by taking averages. 

Unfortunately, the simplicity of the above procedure does not generalize 
easily to the computation of  full one-loop amplitudes. In that case, 
we also have to determine
coefficients of   one-~, two- and three-point scalar integrals. For example,
when a double cut is applied  to determine coefficients of  two-point 
functions, we have to 
account for the fact that parts of these contributions are already contained 
in quadruple and triple cuts. 
Analytic separation of these overlapping contributions
proved to be complicated. 

An efficient algebraic method to separate the 
overlapping contributions was outlined 
in Ref.~\cite{opp}. 
Developing upon this idea, flexible unitarity-based computational 
techniques were suggested in Refs.~\cite{egk,opp2} where it was shown 
how generalized unitarity can be implemented numerically. 
Analytic techniques were further developed in 
Refs.~\cite{Forde:2007mi,Kilgore:2007qr}.
All these papers address primarily 
 cut-constructible parts of  scattering amplitudes.

These developments provide  a polynomial-complexity 
computational algorithm for cut-constructible parts of scattering 
amplitudes. However, techniques for calculating 
rational parts are much less developed.  Three  methods for computing 
rational parts are currently available.
 
In  Refs.~\cite{xyz,binoth} it was suggested to determine  rational
parts of tensor one-loop integrals analytically and  
employ  traditional diagrammatic  methods to obtain scattering amplitudes.  
This leads to  an exponential-, rather than polynomial-complexity
algorithm, negating all
the progress achieved with the determination of the cut-constructible 
part using numerical unitarity techniques.

The other two methods are so far analytic,
but both should in principle be suitable for numerical implementation. The  
so-called bootstrap method sets
up a recursive computation of rational parts~\cite{bootstrap1,bootstrap2} 
similar to 
tree-level unitarity-based recursion relations~\cite{BCFW}. 
However, in its current formulation the bootstrap approach  is not directly 
suitable for numerical implementation since both the 
cut-constructible and rational parts contain   spurious
poles. When the two parts are added together, 
spurious poles cancel. Unitarity-based recursion 
relations for the rational part
can only be constructed once spurious poles are
removed  from the rational part. The procedure to remove these poles 
from the rational part is called cut-completion; it requires 
analytic knowledge of  the cut-constructible part.  
Clearly, for complicated multi-particle processes this
 is a serious disadvantage. 

The other approach  is a variant of    
$D$-dimensional unitarity method~\cite{ABFKM,brittofengUC}. 
%It uses  spinorial technique  with double cuts. 
%The  full $\epsilon$ dependence
%of the amplitude is kept and one can read 
%out $\epsilon$ dependent coefficient of the bubble integrals
%algebraically.  
Since the cut lines are $D$-dimensional, tree scattering 
amplitudes  have to be calculated in $D$-dimensions.
Currently,  this method is restricted to 
analytic applications  for purely gluonic one-loop scattering amplitudes 
with a scalar particle propagating in the loop.  In the following 
sections we describe a computational method based on $D$-dimensional
unitarity that permits straightforward calculation of both cut-constructible 
and rational parts of arbitrary one-loop scattering amplitudes.

\section{One-Loop Amplitudes and Dimensionality of Space-Time}
\begin{figure}[t!]
   \begin{center}
   \leavevmode
   \epsfysize=8.0cm
   \epsffile{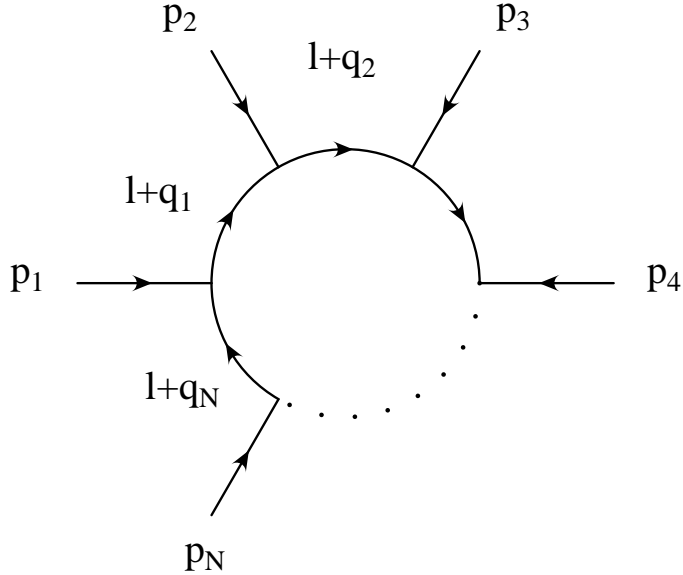}
\end{center}
\caption{The generic $N$-point loop amplitude. }
\label{fig:generic}
\end{figure}

Since  one-loop calculations in quantum field theory lead to 
divergent expressions, we require regularization at intermediate 
stages of the calculation. 
Such  regularization is  accomplished by continuing momenta and 
polarization vectors of unobserved virtual particles 
to $D \ne 4$ dimensions~\cite{'t Hooft:1972fi}.
On the other hand, it is convenient to keep 
momenta and polarization vectors of all external particles
in four dimensions  since this allows
us to define one-loop scattering amplitudes 
through helicities  of external 
particles. Once the  dependence
of a one-loop amplitude on the dimensionality of space-time is 
established,  we interpolate 
to a non-integer number of dimensions $D=4-2\e$. The divergences of one-loop 
amplitudes are regularized  by the  parameter $\e$.

We wish to arrive at a numerical implementation of this procedure.
To this end,  it is crucial to keep the number of dimensions in which 
virtual unobserved particles are allowed to propagate as integer since 
only in this case  loop momenta and polarization sums of unobserved 
particles are fully defined. Therefore,  we determine the dependence 
of one-loop amplitudes on the dimensionality of space-time treating 
the latter as integer and arrive at non-integer values (e.g. $D=4-2\ep$) 
later on, by simple polynomial interpolation.

Any $D$-dimensional cyclic-ordered $N$-particle one-loop 
scattering amplitude (Fig.~\ref{fig:generic})
can be written as
\beq\label{ScatterAmp}
{\cal A_{(D)}}(\{p_i\},\{J_i\})=\int \frac{d^D\, l}{i(\pi)^{D/2}}
\frac{{\cal N}(\{p_i\},\{J_i\};l)}{d_1d_2\cdots d_N}\ ,
\eeq
where $\{p_i\}$ and $\{J_i\}$ are the two 
sets that represent momenta and sources (polarization vectors, 
spinors, etc.) of external particles.
The numerator structure
${\cal N}(\{p_i\},\{J_i\};l)$ depends on the particle content
of the theory.
The denominator is a product of inverse propagators
\beq
d_i=d_i(l)=(l+q_i)^2-m_i^2=\left(l-q_0+\sum_{j=1}^i p_i\right)^2-m_i^2\ ,
\eeq
where the four-vector $q_0$ represents the arbitrary parameterization 
choice of the loop momentum.

The one-loop amplitude can be written as a linear combination 
of master integrals
\beqa\label{decomp}
{\cal A}_{(D)} &=&
\sum_{[i_1|i_5]}\! e_{i_1i_2i_3i_4i_5} 
I^{(D)}_{i_1i_2i_3i_4i_5}+\sum_{[i_1|i_4]}\! d_{i_1i_2i_3i_4} 
I^{(D)}_{i_1i_2i_3i_4} \nonumber \\
&+& \sum_{[i_1|i_3]}\! c_{i_1i_2i_3} I^{(D)}_{i_1i_2i_3}+\sum_{[i_1|i_2]}\! 
b_{i_1i_2} I^{(D)}_{i_1i_2}
+\sum_{[i_1|i_1]}\! a_{i_1} I^{(D)}_{i_1} ,
\eeqa
where we introduced the short-hand notation
$[i_1|i_n] = 1\leq i_1<i_2<\cdots <i_n \leq N$.
The master integrals on the r.h.s. of Eq.~(\ref{decomp}) are defined as
\beq\label{MasterIntegral}
I^{(D)}_{i_1\cdots i_M}=\int \frac{d^Dl}{i(\pi)^{D/2}}
\frac{1}{d_{i_1}\cdots d_{i_M}}\ .
\eeq
The coefficients of  master integrals for this choice of 
basis depend on the number of dimensions $D$ which, in practical calculations
in dimensional regularization,  
needs to be taken as $D=4-2\ep$.  Since we 
aim at numerical implementation of $D$-dimensional unitarity, this 
is inconvenient. We explain below how to change the basis of master 
integrals to make coefficients $D$-independent.

The $D$-dependence of one-loop scattering amplitudes associated with
virtual particles comes from two sources.
When we continue loop momenta and polarization vectors to 
higher-dimensional
space time, the number of spin eigenstates changes.
For example,  massless spin-one particles in $D_s$ dimensions
have $D_s-2$ spin  eigenstates while
spinors in $D_s$ dimensions have $2^{(D_s-2)/2}$ spin eigenstates.
In the latter case, $D_s$ should be even.

The spin density matrix for a massless
spin-one particle with momentum $l$ and polarization vectors
$e_{\mu}^{(i)}$ is given by
\be
\sum \limits_{i=1}^{D_s-2} e_{\mu}^{(i)}(l)e_{\nu}^{(i)}(l)=
-g_{\mu\nu}^{(D_s)}+\frac{l_{\mu}b_{\nu}+b_{\mu}l_{\nu}}{l\cdot b},
\ee
where $b_{\mu}$ is an arbitrary light-cone gauge vector associated
with a particular choice of polarization vectors. Similarly,
the spin density matrix for a fermion with momentum
$l$ and mass $m$ is given by
%and Dirac spinor $u^{(i)}(l)$ is given by
\be
\sum_{i=1}^{2^{(D_s-2)/2}} u^{(i)}(l)\overline{u}^{(i)}(l)
=\s{l}+m=\sum_{\mu=1}^{D} l_{\mu}\gamma^{\mu}+m\ .
\ee
While, as we see from these examples,  the
number of spin eigenstates depends explicitly on the space-time 
dimensionality,
the loop-momentum $l$ itself has implicit $D$-dependence. We can define
the loop momentum as a $D$-dimensional vector, with the requirement
$D\leq D_s$~\cite{Bern:2002zk}.
We now extend the notion of dimensional dependence of the
one-loop scattering amplitude in Eq.~(\ref{ScatterAmp}) by taking 
the sources of all
unobserved particles in $D_s$-dimensional space-time
\be
{\cal A}_{(D,D_s)}(\{p_i\},\{J_i\}) =\int \frac{{ d}^D\, l}{i(\pi)^{D/2}}
%\int \frac{{\rm d}^Dl}{(2\pi)^D}
%{\cal A}_{(D_s)}(\{p_i\},\{J_i\};l)-
%\int \frac{{\rm d}^Dl}{(2\pi)^D}
\frac{{\cal N}^{(D_s)}(\{p_i \},\{J_i\};l )}{d_1d_2\cdots d_N}.
\label{eq1}
\ee
The numerator function
${\cal N}^{(D_s)}(\{p_i\},\{J_i \}; l)$ depends explicitly on $D_s$
through the number of spin eigenstates of virtual particles.
However, the dependence of the numerator function on the
loop momentum dimensionality $D$ emerges in a peculiar way.
Since external particles are kept in
four dimensions, the dependence of the numerator function
 on $D-4$ components
of the loop momentum $l$ appears only through its dependence on $l^2$.
Specifically
\be
l^2=\overline{l}^2-\widetilde{l}^2=l_1^2-l_2^2-l_3^2-l_4^2-\sum_{i=5}^{D} 
l_i^2\ ,
\label{eq0}
\ee
where $\overline{l}$ and $\widetilde{l}$ denote 
four- and $(D-4)$-dimensional
components of the vector $l$.
It is apparent from Eq.~(\ref{eq0}) that 
 there is no preferred direction in the $(D-4)$-dimensional
subspace  of the $D$-dimensional loop momentum space.

A simple, but important observation is that  in one-loop calculations,
the dependence of  scattering amplitudes
on $D_s$ is {\it linear}. This happens because, for such dependence to
appear, we need to have a closed loop of contracted
metric tensors and/or Dirac matrices coming from  vertices and propagators.
Since only a single loop can appear in one-loop calculations, we find
\be
{\cal N}^{(D_s)} (l) = {\cal N}_0(l) + (D_s - 4) {\cal N}_1(l).
\label{eq2}
\ee
We emphasize that there is no explicit dependence on either $D_s$ or
$D$ in functions ${\cal N}_{0,1}$.

For numerical calculations we need to
separate the two functions ${\cal N}_{0,1}$. To do so, we compute the left
hand side of Eq.~(\ref{eq2}) for $D_s = D_1$ and $D_s =D_2 $ and, after 
taking
appropriate linear combinations, obtain
\ba
\label{eq3}
{\cal N}_0(l) &=& \frac{(D_2-4){\cal N}^{(D_1)}(l) - (D_1-4){\cal 
N}^{(D_2)}(l)}{D_2-D_1}, \nn
{\cal N}_1(l) &=& \frac{{\cal N}^{(D_1)}(l) - {\cal N}^{(D_2)}(l)}{D_2-D_1}.
\ea
Because both $D_1$ and $D_2$ are integers,  
amplitudes are numerically well-defined. We will comment more on possible
choices of $D_{1,2}$ in the forthcoming sections; here suffice it to say 
that
if fermions are present in the loop, we have
to choose {\it even} $D_1$ and $D_2$.
%so that, similar to
%four-dimensional case,  the Clifford algebras have only one equivalent
%class of irreducible representations in terms of $n \times n$ matrices
%where $n = 2^{D_s/2}$.

Having established the $D_s$-dependence of the amplitude, we
discuss analytic continuation for sources of unobserved
particles.  We can interpolate $D_s$
either to $D_s\to 4-2\e$ (the t'Hooft-Veltman (HV) scheme)~\cite{'t Hooft:1972fi}
or to $D_s\to 4$ (the four-dimensional helicity (FDH) scheme)~\cite{Bern:2002zk}.
The latter scheme is of particular interest in supersymmetric (SUSY)
calculations since
all SUSY Ward identities are preserved. We see from Eq.~(\ref{eq2})
that the difference
between the two schemes is simply $-2\e{\cal N}_1$.
%\\ \WG{Is the scheme difference simply picking up the UV-pole, which 
%per definition is% proportional to Born?} \\

We now substitute Eq.(~\ref{eq3}) into Eq.~(\ref{eq1}). Upon doing so,
we obtain  explicit expressions for one-loop amplitudes in HV and FDH
schemes. We derive
%(\{p_i\},\{J_i\})
\ba
{\cal A}^{\rm FDH} &=&
\left(\frac{D_2-4}{D_2-D_1}\right){\cal A}_{(D, D_s =D_1)}
-\left(\frac{D_1-4}{D_2-D_1}\right){\cal A}_{(D, D_s =D_2)}, \nonumber \\
{\cal A}^{\rm HV} &=& {\cal A}^{\rm FDH}
-\left(\frac{2\e}{D_2-D_1}\right)\left({\cal A}_{(D, D_s =D_1)}
-{\cal A}_{(D, D_s =D_2)}\right).
\label{eq2.4}
\ea
We emphasize that $D_s=D_{1,2}$ amplitudes on the r.h.s. of 
Eq.~(\ref{eq2.4})
are conventional one-loop  scattering amplitudes  whose
numerator functions are computed in higher-dimensional
space-time, i.e. all internal metric tensors and Dirac gamma matrices are in
integer $D_s = D_{1,2}$ dimensions.
The loop integration is in $D\leq D_s$ dimensions.
It is important that explicit dependence on the regularization
parameter $\ep = (4-D)/2$
%generated by spin density matrices  
is not present in these amplitudes.  
For this reason, Eq.~(\ref{eq2.4})
renders itself to straightforward numerical implementation.
In particular,  numerical implementation
of $D_s$-dimensional unitarity cuts is now straightforward,
 as cut internal lines possess well-defined spin density matrices.
The $D_s$-dimensional unitarity cuts, that we discuss in detail in the
next section,  decompose
amplitudes into a linear combination of master integrals, see 
Eq.~(\ref{decomp}).
We can choose the basis of master integrals in such a way that
no explicit $D$-dependence in the coefficients appears.
%thereby making the $D$-dimensional dependence of the coefficients explicit.
Only after the reduction to master integrals is established,
we  continue
the space-time dimension associated with the loop momentum
to $D\to 4-2\e$,
% This generates proper $D$-dimensional
% dependence of the coefficients and
thereby regularizing the master integrals.

\section{$D_s$-dimensional Unitarity Cuts}

The amplitudes on the r.h.s. of Eq.~(\ref{eq2.4}) are most efficiently
calculated by using generalized unitarity applied in $D_s$  dimensions.
Since many aspects of the calculation in this case are the same as in
four-dimensional generalized unitarity, we focus on new features that 
appear for $D_s>4$. Our discussion  follows Ref.~\cite{egk}
which details the $D_s=4$ case.

The important  issue is the set  of master integrals that we have to deal
with  when applying $D_s$-dimensional unitarity cuts.
Cutting a propagator $i$ requires finding the loop momentum $l$ such that
equation  $d_i(l)=0$ is satisfied. Therefore, each cut imposes
one constraint on the loop momentum. In four dimensions, where loop momentum
has only four components, one can cut at most
four propagators without over-constraining  the system of equations. This
leads to the conclusion that four-point integrand functions in 
four-dimensions
are required whereas five-point and higher-point functions aren't.

When we compute an  amplitude in $D_s > 4$ dimensions, we may also use 
the additional
$D-4$ components of the loop momenta
to set more inverse propagators to zero. However, since all external
momenta are
four-dimensional, additional components of the loop momentum enter
all propagators in a particular combination
\be
s_e^2  = -\sum \limits_{i=5}^{D} (l \cdot n_i)^2= -\sum 
\limits_{i=5}^{D} (\widetilde{l} \cdot n_i)^2,
\label{lne}
\ee
where $n_{i>4}$ are orthonormal
basis vectors ($n_i\cdot n_j=\delta_{ij}$) of the $(D-4)$-dimensional 
sub-space embedded in a $D$-dimensional space.
Therefore, when we move from four to $D_s$ dimensions, at most {\it five}
inverse propagators can be set to zero.
Therefore, for $D_s > 4$, the highest-point master integral that should be 
included into the master integrals basis is a five-point function.
Hence the integrand of the $N$-particle amplitude in 
Eq.~(\ref{ScatterAmp}) can be
parameterized as
\beqa\label{partial}
\frac{{\cal N}^{(D_s)}(l)}{d_1d_2\cdots d_N} &=&
\sum_{[i_1|i_5]} 
\frac{\overline{e}^{(D_s)}_{i_1i_2i_3i_4i_5}(l)}{d_{i_1}d_{i_2}d_{i_3}d_{i_4}d_{i_5}}
+\sum_{[i_1|i_4]} 
\frac{\overline{d}^{(D_s)}_{i_1i_2i_3i_4}(l)}{d_{i_1}d_{i_2}d_{i_3}d_{i_4}}
\nonumber \\
&+&\sum_{[i_1|i_3]} 
\frac{\overline{c}^{(D_s)}_{i_1i_2i_3}(l)}{d_{i_1}d_{i_2}d_{i_3}}
+\sum_{[i_1|i_2]} \frac{\overline{b}^{(D_s)}_{i_1i_2}(l)}{d_{i_1}d_{i_2}}
+\sum_{[i_1|i_1]} \frac{\overline{a}^{(D_s)}_{i_1}(l)}{d_{i_1}}\,.
\eeqa
where the dependence on the external momenta and sources are suppressed.
 From four-dimensional unitarity we know that computation of each cut
of the scattering amplitude is simplified if convenient parameterization
of the residue is chosen. We now discuss how these parameterizations change
when $D_s$-dimensional unitarity cuts are considered.

\subsection{Pentuple residue}

To calculate the pentuple residue, we choose momentum $l$ such that
five inverse propagators in Eq.~(\ref{partial}) vanish. We define
\be
\overline{e}^{(D_s)}_{ijkmn}(l_{ijkmn} ) = {\rm Res}_{ijkmn} \left (
\frac{{\cal N}^{(D_s)}(l)}{d_1\cdots d_N} \right ).
\ee
The momentum $l_{ijkmn}$ satisfies the following
set of equations $d_i(l_{ijkmn}) =\cdots=d_n(l_{ijkmn}) = 0$.
The solution is given by
\be
l^{\mu}_{ijkmn}=V_5^{\mu}
+\sqrt{\frac{-V_5^2 + m_n^2}{\alpha_5^2+\cdots+\alpha_D^2}}\left(\sum_{h=5}^D 
\alpha_h n_h^{\mu}\right)\ ,
\ee
where $m_n$ is the mass in the propagator $d_n$ which is chosen to be as
$d_n = l^2 - m_n^2$ by adjusting the reference vector $q_0$. The  
parameters $\alpha_h$ can be chosen freely. The four-dimensional 
vector $V_5^{\mu}$ depends only on external momenta and propagator masses.
It is explicitly constructed
using the Vermaseren-van Neerven basis as outlined in Ref.~\cite{egk}.
The $D-4$ components of the vector $l_{ijkmn}$ are necessarily 
non-vanishing;
for simplicity we may choose $l_{ijkmn}$ to be five-dimensional,
independent of $D_s$. We will see below that this is
sufficient to determine pentuple residue.

To restrict the functional form of the pentuple residue 
$\overline{e}_{ijkmn}(l)$
we apply the same  reasoning as in four-dimensional unitarity
case,  supplemented with the requirement that
$\overline{e}^{(D_s)}_{ijkmn}(l)$ depends only on even powers of $s_e$; 
this requirement
is a necessary consequence of the discussion around Eq.~(\ref{eq0}).
These considerations  lead to the conclusion
that the pentuple residue is independent of the loop momentum
\be
\overline{e}^{(D_s)}_{ijkmn}(l) = e^{(D_s,(0))}_{ijkmn}.
\ee

To calculate $e^{(0)}$ in the FDH scheme, we employ Eq.~(\ref{eq3}) and 
obtain
\be
e^{(0),\rm FDH}_{ijkmn} =
\left(\frac{D_2-4}{D_2-D_1}\right)
{\Res}_{ijkmn} \left (
\frac{{\cal N}^{(D_1)}(l)}{d_1\cdots d_N}
\right )
- \left(\frac{D_1-4}{D_2-D_1}\right)
{\Res}_{ijkmn} \left (
\frac{{\cal N}^{(D_2)}(l)}{d_1\cdots d_N}
\right ).
\label{eq5}
\ee
The calculation of the residues of the amplitude on the r.h.s. of Eq.~(\ref{eq5}),
is simplified by their factorization into products of tree amplitudes
\ba
\label{eq51}
&& {\Res}_{ijkmn} \left ( \frac{{\cal N}^{(D_s)}(l)}{d_1\cdots d_N}
\right )
= \sum   {\cal M}(l_i;p_{i+1},\ldots ,p_j,-l_j )
\times {\cal M}(l_j;p_{j+1},\ldots ,p_k;-l_k)
\\
&&
\times {\cal M}(l_k;p_{k+1},\ldots ,p_m;-l_m)
\times {\cal M}(l_m;p_{m+1},\ldots ,p_n;-l_n)
\times {\cal M}(l_n;p_{n+1}...,p_i;-l_i). \nonumber
\ea
Here,  the summation is over all different quantum numbers of the cut lines.
In particular, we have to sum  over polarization vectors of the cut lines. 
This
generates explicit $D_s$ dependence of the residue, as described in the
previous section. Note that the complex momenta $l_h^{\mu}=l^{\mu}+q_h^{\mu}$
are on-shell due to the unitarity constraint $d_h=0$.

\subsection{Quadrupole residue}

The construction of the quadrupole residue follows the discussion of the
previous subsection and generalizes the four-dimensional case studied in~\cite{opp,egk}.  
We define
\be
\overline{d}_{ijkn}^{(D_s)}(l) = \Res_{ijkn} \left (
\frac{{\cal N}^{(D_s)}(l)}{d_1\cdots d_N}
- \sum  \limits_{[i_1|i_5]}^{} \frac{e^{(D_s,(0))
}_{i_1i_2i_3i_4i_5}}{d_{i_1} d_{i_2} d_{i_3} d_{i_4} d_{i_5}} \right ),
\label{eq6}
\ee
where the last term in the r.h.s. is the necessary subtraction of
the pentuple cut contribution.  We now specialize to the FDH scheme.
In this case, the most general parameterization of the quadrupole
cut is given by
\be
\overline{d}_{ijkn}^{\rm FDH}(l) = d^{(0)}_{ijkn} + d^{(1)}_{ijkn} s_1
+  ( d^{(2)}_{ijkn} + d^{(3)}_{ijkn} s_1 )s_e^2
+ d^{(4)}_{ijkn}s_e^4,
\label{eq7}
\ee
where $s_1=l \cdot n_1$.
We used the fact that, in renormalizable quantum field theories,
the highest rank of a tensor integral that may
contribute to a quadrupole residue is four and that only even powers of
$s_e$ can appear on the r.h.s of Eq.~(\ref{eq7}).
The solution of the unitarity constraint is given by
\be\label{solution4}
l^{\mu}_{ijkn}=V_4^{\mu}+\sqrt{\frac{-V_4^2 + m_n^2}{\alpha_1^2+\alpha_5^2+\cdots+\alpha_D^2}}
\left(\alpha_1 n_1^{\mu}+\sum_{h=5}^D \alpha_h n_h^{\mu}\right)\ .
\ee
The vector $V_4$ is defined in the space spanned by the three independent 
inflow momenta $\{k_i, k_j, k_k\}$
and $n_1$ is the  unit vector that describes the  one-dimensional
``transverse space'', i.e. $k_i\cdot n_1=0$, $k_j\cdot n_1=0$, $k_k\cdot 
n_1=0$, $n_i\cdot n_j=\delta_{ij}$~\cite{egk}.

To determine the momentum-independent  coefficients 
$d^{(0,1,\ldots,4)}_{ijkn}$ in Eq.~(\ref{eq7}), we compute
the quadrupole residue of the one-loop scattering amplitude for different
values of the loop momentum $l$ that satisfies the unitarity constraint.
This entails choosing different values for parameters $\alpha_i$
in Eq.~(\ref{solution4}).
As a first step, we may choose $l$ to be a four-dimensional 
vector embedded in $D_s$-dimensional space ($\alpha_{i\geq 5}=0$). Then 
$s_e = 0$ and the
parameterization of the residue in Eq.~(\ref{eq7}) becomes
identical to a four-dimensional case.  Standard
manipulations described in Refs.~{\cite{opp,egk}} then allow
us to find $d^{(0)}_{ijkn}$ and $d^{(1)}_{ijkn}$.  To determine the
remaining coefficients, we consider loop momenta in dimensions $D>4$
such that $d_i=d_j=d_k=d_m=0$ but $s_e \ne 0$. We accomplish this 
by adjusting the value of $\alpha_5$. 
By choosing  appropriate 
loop momenta, we
determine all the remaining three coefficients of the quadrupole residue 
by solving a
linear system of equations.

\subsection{Triple , double and single-line   cuts}

Calculation of triple, double and single cuts proceeds in full analogy
with what has been described for pentuple and quadrupole cuts.
The only modification concerns parameterization of residues. 
%We display the necessary formulas below.

The  general parameterization of a triple cut in the FDH scheme is given by
\beqa
\overline{c}^{\rm 
FDH}_{ijk}(l)&=&c_{ijk}^{(0)}+c_{ijk}^{(1)}s_1+c_{ijk}^{(2)}s_2+c_{ijk}^{(3)}(s_1^2-s_2^2)
+s_1s_2(c_{ijk}^{(4)}+c_{ijk}^{(5)}s_1+c_{ijk}^{(6)}s_2)\nn
&&+ c^{(7)}_{ijk} \;s_1 \; s_e^2
 + c^{(8)}_{ijk} \;s_2 \;s_e^2
 + c^{(9)}_{ijk} s_e^2,
\eeqa
where $s_1=l \cdot n_1$ and  $s_2=l \cdot n_2$.
The solution of the unitarity constraint is given by
\be\label{solution3}
l^{\mu}_{ijk}=V_3^{\mu}+\sqrt{\frac{-V_3^2+m_k^2}{\alpha_1^2+\alpha_2^2+\alpha_5^2+\cdots+\alpha_D^2}}
\left(\alpha_1 n_1^{\mu}+\alpha_2 n_2^{\mu}+\sum_{h=5}^D \alpha_h n_h^{\mu}\right)\ .
\ee
The vector $V_3$ is defined in the space spanned by the two independent 
inflow momenta $\{k_i, k_j\}$
and $n_{1,2}$ are orthonormal vectors that describe the 
two-dimensional
``transverse space'', i.e. $k_i\cdot n_{1,2}=0$, $k_j\cdot n_{1,2}=0$, 
$n_i\cdot n_j=\delta_{ij}$~\cite{egk}.

The  general parameterization of a double cut is given by
\beqa
\overline{b}^{\rm 
FDH}_{ij}(l)&=&b_{ij}^{(0)}+b_{ij}^{(1)}s_1+b_{ij}^{(2)}s_2+b_{ij}^{(3)}s_3
+b_{ij}^{(4)}(s_1^2-s_3^2)+b_{ij}^{(5)}(s_2^2-s_3^2)
+b_{ij}^{(8)}s_2s_3\nn
&+& b_{ij}^{(6)}s_1s_2+b_{ij}^{(7)}s_1s_3
+ b_{ij}^{(9)} s_e^2,
\eeqa
where $s_1=l \cdot n_1$, $s_2=l \cdot n_2$ and $s_3=l \cdot n_3$.
The solution of the unitarity constraint is given by
\be\label{solution2}
l^{\mu}_{ij}=V_2^{\mu}+\sqrt{\frac{-V_2^2+m_j^2}{\alpha_1^2+\alpha_2^2+\alpha_3^2+\alpha_5^2+\cdots+\alpha_D^2}}
\left(\alpha_1 n_1^{\mu}+\alpha_2 n_2^{\mu}+\alpha_3 n_3^{\mu}+\sum_{h=5}^D 
\alpha_h n_h^{\mu}\right)\ .
\ee
The vector $V_2$ is proportional to the inflow momentum $\{k_i\}$
and $n_{1,2,3}$ are orthonormal vectors that describe the
three-dimensional
``transverse space'', i.e. $k_i\cdot n_{1,2,3}=0$, $n_i\cdot 
n_j=\delta_{ij}$~\cite{egk}.
Finally, we note that the
parameterization
of a single-line residue is the same as in the four-dimensional case
described in Ref.~\cite{egk}.

\subsection{One-loop amplitudes and master integrals}

Following the strategy outlined in previous subsections, we obtain
pentuple, quadrupole, triple, double and single-line residues.
These residues give us coefficients
of master integrals through which the one-loop amplitude is expressed.

Because we have more coefficients in our parameterization of residues,
than in the four-dimensional case,  we  end up with a larger number of
master integrals. The new master integrals include  the five-point function
and various $s_e$-dependent terms that appear in quadrupole, triple and
double residues. We will comment on the fate of the five-point function
shortly. 

Consider now the $s_e$-dependent master integrals. Some of them contain
scalar products $l\cdot n_{1,2}$ and vanish upon angular integration
over $l$.
Neglecting these spurious terms,
we have to deal with  four additional master integrals. We can rewrite
those integrals through conventional four-, three- and two-point
functions in higher-dimensional space-time. We find
\ba
\label{eq3.1}
&& \int \frac{{\rm d}^D l}{(i\pi)^{D/2}}
\frac{s_e^2}{d_{i_1} d_{i_2} d_{i_3} d_{i_4}}
= -\frac{D-4}{2}I^{D+2}_{i_1 i_2 i_3 i_4}, \nonumber \\
&& \int \frac{{\rm d}^D l}{(i\pi)^{D/2}}
\frac{s_e^4}{d_{i_1} d_{i_2} d_{i_3} d_{i_4}}
= \frac{(D-2)(D-4)}{4}I^{D+4}_{i_1 i_2 i_3 i_4},  \\
&& \int \frac{{\rm d}^D l}{(i\pi)^{D/2}}
\frac{s_e^2}{d_{i_1} d_{i_2} d_{i_3}}
= -\frac{(D-4)}{2}I^{D+2}_{i_1 i_2 i_3}, \nonumber \\
&& \int \frac{{\rm d}^D l}{(i\pi)^{D/2}}
\frac{s_e^2}{d_{i_1} d_{i_2}}
= -\frac{(D-4)}{2}I^{D+2}_{i_1 i_2}. \nonumber
\ea

Using Eq.~(\ref{eq3.1}), we arrive at the following representation of the
scattering amplitude
\beqa\label{MasterDecomp}
&& {\cal A}_{(D)}=
\sum_{[i_1|i_5]} e^{(0)}_{i_1i_2i_3i_4i_5}\
I^{(D)}_{i_1i_2i_3i_4i_5} \nn
&&+\sum_{[i_1|i_4]} \left(
d^{(0)}_{i_1i_2i_3i_4}\ I^{(D)}_{i_1i_2i_3i_4}
-\frac{D-4}{2}\, d^{(2)}_{i_1i_2i_3i_4}\ I^{(D+2)}_{i_1i_2i_3i_4}
%\right.\nn &&\phantom {\sum_{[i_1|i_4]}}\left.
+\frac{(D-4)(D-2)}{4}\, d^{(4)}_{i_1i_2i_3i_4}\ I^{(D+4)}_{i_1i_2i_3i_4}
\right)\nonumber \\
&&+\sum_{[i_1|i_3]} \left(c^{(0)}_{i_1i_2i_3}\
I^{(D)}_{i_1i_2i_3}-\frac{D-4}{2}c^{(9)}_{i_1i_2i_3}\
I^{(D+2)}_{i_1i_2i_3} \right)\nn
&& +\sum_{[i_1|i_2]} \left( b^{(0)}_{i_1i_2}\
I^{(D)}_{i_1i_2}-\frac{D-4}{2}b^{(9)}_{i_1i_2}\
I^{(D+2)}_{i_1i_2}\right)
+\sum_{i_1=1}^{N} a^{(0)}_{i_1}\ I^{(D)}_{i_1}\ .
\eeqa
We emphasize
that the explicit
$D$-dependence on the r.h.s. of Eq.~(\ref{MasterDecomp})
is the consequence of our choice
of the  basis for master integrals in Eq.~(\ref{eq3.1}).

We note that the above decomposition is valid for any value of $D$. We
can now interpolate the loop integration dimension $D$
to $D\to 4-2\e$. The extended basis of master integrals that we employ
provides a clear
separation between cut-constructible and rational parts of the amplitude.
The cut-constructible part is
given by the integrals in $D$-dimensions in Eq.~(\ref{MasterDecomp}), while
 the rational part is  given by the integrals in $D+2$ and $D+4$ dimensions.
However, it is possible to use smaller  basis of master
integrals by rewriting integrals $\{I_{i_1i_2i_3i_4}^{(D+4)},
I_{i_1i_2i_3i_4}^{(D+2)}, I_{i_1i_2i_3}^{(D+2)}, I_{i_1i_2}^{(D+2)}\}$
in terms of $\{I_{i_1i_2i_3i_4}^{(D)}, I_{i_1i_2i_3}^{(D)}, I_{i_1i_2}^{(D)}\}$
using the integration-by-parts techniques.

Since we are interested in NLO computations,
we only need to consider
the limit $\ep \to 0$ in Eq.~(\ref{MasterDecomp})
 and neglect contributions of order $\e$. This leads to certain
simplifications.
First, in this limit,  we can re-write the scalar 5-point master 
integral as a linear combination
of four-point master integrals up to ${\cal O}(\ep)$ terms. If we employ
this fact in Eq.~(\ref{MasterDecomp}), we obtain
\beq\label{5-pointRedux}
\lim_{D \to 4-2\ep} \; \left(
\sum_{[i_1|i_5]} e^{(0)}_{i_1\cdots i_5}\
I^{(D)}_{i_1\cdots i_5}
+\sum_{[i_1|i_4]}
d^{(0)}_{i_1\cdots i_4}\ I^{(D)}_{i_1\cdots i_4} \right)
=\sum_{[i_1|i_4]}
\tilde d^{(0)}_{i_1\cdots i_4}\ I^{(4-2\ep)}_{i_1\cdots i_4}
+{\cal O}(\ep)\ .
\eeq
Note that since {\it scalar} five-point function is cut-constructible,
it does not contribute to the rational part.
Second, since  $\lim_{D \to 4}(D-4)\times I_{i_1i_2i_3i_4}^{(D+2)}= 0$,
the 6-dimensional 4-point integral can be neglected.
Finally, the 6-dimensional 2-point, 6-dimensional 3-point and
8-dimensional 4-point integrals are all ultraviolet divergent and
produce contributions of order $\ep^{-1}$.
Therefore, factors $(D-4)$ that multiply these integrals in
Eq.~(\ref{MasterDecomp}) 
pick up  divergent terms  and produce $\ep$-independent,
finite contributions
\beqa
&& \lim_{D \to 4}\;\;
\frac{(D-4)(D-2)}{4}\ I^{(D+4)}_{i_1i_2i_3i_4}=-\frac{1}{3},
 \nn
&& \lim_{D \to 4}\;\;
\frac{(D-4)}{2}\ I^{(D+2)}_{i_1i_2i_3}=\frac{1}{2}, \\
&& \lim_{D \to 4}\;\;
\frac{(D-4)}{2}\
I^{(D+2)}_{i_1i_2}=-\frac{m_{i_1}^2 + m_{i_2}^2}{2}
+ \frac{1}{6}\left(q_{i_1}-q_{i_2}\right)^2. \nonumber
\eeqa

Combining everything together, we can write any one-loop amplitude
up to terms of order $\ep$
as a linear combination of cut-constructible and rational parts.
\be
{\cal A}_N={\cal A}^{CC}_N+R_N.
\ee
The expression for the cut-constructible part reads
\be
{\cal A}^{CC}_N = \sum_{[i_1|i_4]}
\tilde d^{(0)}_{i_1i_2i_3i_4}\ I^{(4-2\e)}_{i_1i_2i_3i_4}
+\sum_{[i_1|i_3]} c^{(0)}_{i_1i_2i_3}\ I^{(4-2\e)}_{i_1i_2i_3}
+\sum_{[i_1|i_2]} b^{(0)}_{i_1i_2}\ I^{(4-2\e)}_{i_1i_2}
+\sum_{i_1=1}^{N} a^{(0)}_{i_1}\ I^{(4-2\e)}_{i_1},
\ee
where coefficients $\tilde d^{(0)}_{i_1\cdots i_4}$ are implicitly defined
in Eq.~(\ref{5-pointRedux}). For the rational part, we obtain
\be
R_N= -\sum_{[i_1|i_4]}
\frac{d^{(4)}_{i_1i_2i_3i_4}}{3}
-\sum_{[i_1|i_3]} \frac{c^{(9)}_{i_1i_2i_3}}{2}
-\sum_{[i_1|i_2]}
\left ( \frac{(q_{i_1}-q_{i_2})^2}{6}-\frac{m_{i_1}^2 + m_{i_2}^2}{2}
\right )\
b^{(9)}_{i_1i_2}.
\ee
Finally, we point out that various master integrals required
for one-loop computations can be found
in  Ref.~\cite{rep}.

\section{Gluon scattering amplitudes in QCD}

We applied the method described in the previous sections to compute 
gluon scattering amplitudes in QCD. It is well-known that these amplitudes
can be represented as linear combinations of simpler objects, 
called  color-ordered sub-amplitudes.  For example, the tree
amplitude for $n$-gluon scattering reads~~\cite{Berends:1987cv,Mangano:1987xk} 
\be
{\cal A}_{n}^{\rm tree}(\{ p_i,\lambda_i,a_i \}) 
= g^{n-2} \sum \limits_{\sigma \in S_n/Z_n}
{\rm Tr}\left ( T^{a_{\sigma(1)}}\cdots T^{a_{\sigma(n)}} \right )
{\cal A}_{n}^{\rm tree}\left ( p^{\lambda_{\sigma(1)}}_{\sigma(1)},\ldots ,
p^{\lambda_{\sigma(n)}}_{\sigma(n)} \right ),
\label{eq4.0}
\ee
where $p_i, \lambda_i, a_i$ stand for momenta, helicities and color indices 
of external gluons,  $g$ is the coupling constant and 
$T^a$ are generators of the ${\rm SU}(N_c)$ color algebra normalized
as ${\rm Tr}(T^a T^b) = \delta^{ab}$. The sum in Eq.~(\ref{eq4.0})
runs over $(n-1)!$  non-cyclic permutations of the set $\{1,\ldots ,n\}$. 
Amplitudes ${\cal A}_{n}^{\rm tree}\left ( p^{\lambda_{\sigma(1)}}_{\sigma(1)},\ldots ,
p^{\lambda_{\sigma(n)}}_{\sigma(n)} \right )$ are color-ordered tree 
sub-amplitudes.

One-loop amplitudes can be decomposed in a similar, but slightly more 
complicated fashion~\cite{Bern:1990ux}. Considering one-loop amplitudes 
for $n$-gluon scattering, where all 
internal particles are also gluons, we write
\be
{\cal A}_n\left ( \{ p_i, \lambda_i, a_i \} \right ) = 
g^n 
\sum \limits_{c = 1}^{[n/2]+1}
\sum \limits_{\sigma \in S_n/S_{n;c}}
{\rm Gr}_{n;c}(\sigma) {\cal A}_{n,c}(\sigma ),
\label{eq4.0a}
\ee
 where 
\be
{\rm Gr}_{n;1}(\sigma) = 
N_c {\rm Tr}\left ( T^{a_{\sigma(1)}}\cdots T^{a_{\sigma(n)}} \right ),
\label{eq4.0b}
\ee
and 
\be
{\rm Gr}_{n;c}(\sigma) = 
{\rm Tr}\left ( T^{a_{\sigma(1)}}\cdots T^{a_{\sigma(c-1)}} \right )
{\rm Tr}\left ( T^{a_{\sigma(c)}}\cdots T^{a_{\sigma(n)}} \right ).
\label{eq4.0c}
\ee
In Eq.~(\ref{eq4.0a}) $[x]$ is the largest integer number smaller than 
or equal  to $x$ and $S_{n,c}$ are subsets of $S_n$ that leave  
the double trace in Eq.~(\ref{eq4.0c}) invariant.

Since sub-amplitudes possess a number of symmetries
under cyclic permutations of external particles and parity, 
not all ${\cal A}_{n,1}$
amplitudes are independent. Moreover, all ${\cal A}_{n,c>1}$ amplitudes can be written
as linear combinations of ${\cal A}_{n,1}$ amplitudes. Because of that,
there are four independent helicity sub-amplitudes for four- and 
five-gluon scattering and eight independent helicity sub-amplitudes for six-gluon 
scattering.  Scattering amplitudes for four, five and 
six gluons are available 
in the literature; this  allows us to check 
the validity of our method. 

The method described in the previous section is amenable to straightforward 
numerical implementation. We construct a list of all possible cuts of 
a given sub-amplitude. Those cuts that correspond to 
two-point functions for light-like incoming momenta 
are discarded since corresponding master integrals vanish in dimensional 
regularization.

Each cut in the list is computed as a product of tree 
amplitudes which are obtained using Berends-Giele recurrence 
relations~\cite{Berends:1987me}. The recurrence relations themselves do not change 
when we construct gluon scattering amplitudes in higher-dimensional space-time.
However, polarization vectors for cut gluon lines have to be extended.
To discuss this extension explicitly, we now choose specific 
values for space-time dimensionalities $D_{1,2}$ in Eq.~(\ref{eq2.4}).
Since we deal with pure gluonic amplitudes, we may consider $D_1 = 5$ 
and $D_2 = 6$. We obtain
\be
{\cal A}^{\rm FDH} = 2 {\cal A}_{(D,D_s=5)} - {\cal A}_{(D,D_s=6)}.
\label{eq29}
\ee
Computation of residues discussed in the previous section 
requires $s_e \ne 0$. This can be easily accomplished 
 by allowing the projection 
of the loop momentum on the fifth direction to be non-vanishing, while always keeping  $l \cdot n_6 = 0$, even for $D_s = 6$. Hence, for computing 
residues of gluon amplitudes we have to consider a few cases of how four-
and five-dimensional loop momentum can be embedded into five- and six-dimensional space-time.  For the sake of clarity, we describe those cases separately.

For $D_s=5$ and four-dimensional loop momentum, we have $l \cdot n_5 = 0$. 
This allows us to write 
\be
\sum \limits_{i=1}^{3} e_{i}^{\mu} e_i^{\nu} = 
\left \{ \begin{array}{cc}
\rho^{\mu \nu}(l,\eta), & \mu,\nu \in 4\ {\rm dim}; \\
-n_5^{\mu} n_5^{\nu}, & \mu=5, \nu=5;\\ 
0, &~~~{\rm otherwise},
\end{array} \right.
\label{eq5a4}
\ee
where 
\be
\rho^{\mu \nu}(l,\eta) = -g^{\mu \nu} + 
\frac{(l^\mu \eta^\nu + l^\nu \eta^\nu)}{l \cdot \eta},
\label{eq4a4}
\ee
and  $\eta$ is a four-dimensional light-cone vector such that 
$l \cdot \eta \ne 0$.

For $D_s=6$ and four-dimensional loop momentum, we have $l \cdot n_5 = l \cdot n_6 = 0$.
We then choose the following expression for gluon density matrix
\be
\sum \limits_{i=1}^{4} e_{i}^{\mu} e_i^{\nu} = 
\left \{ \begin{array}{cc}
\rho^{\mu \nu}(l,\eta), & \mu,\nu \in 4\ {\rm dim}; \\
-n_5^{\mu} n_5^{\nu}, & \mu=5, \nu=5;\\ 
- n_6^{\mu} n_6^{\nu}, & \mu = 6, \nu = 6;\\
0, &~~~~{\rm otherwise}.
\end{array} \right.
\label{eq6a4}
\ee

For $D_s=5$ and five-dimensional loop momentum, 
we write $l^\mu = \overline{l}^{\mu} + \beta n_5^{\mu}$, 
where $\overline{l} \cdot n_5 = 0$. Also, $l^2 = 0$, 
but $\overline{l}^2 = -\beta^2 \ne 0$.
Then, we  construct polarization vectors by taking them 
to be in the three-dimensional subspace of a five-dimensional space, 
which is orthogonal to $\overline{l}$ and $n_5$. 
Specifically, we have 
\be
\sum \limits_{i=1}^{3} e_{i}^{\mu} e_i^{\nu} = 
\left \{ \begin{array}{cc}
-\omega^{\mu \nu} (\overline{l}); & \mu,\nu \in 4\ {\rm dim};\\
0, & {\rm otherwise}, 
\end{array} \right.
\label{eq5a5}
\ee
where $\omega^{\mu \nu} = - g^{\mu \nu} + \overline{l}^{\mu} \overline{l}^{\nu}/\overline{l}^2$, 
with all indices restricted to four-dimensions.
From a four-dimensional viewpoint, Eq.~(\ref{eq5a5}) is a density matrix 
of a spin-one particle with the mass $\overline{l}^2$.

For $D_s=6$ and five-dimensional loop momentum, 
we take $l = \overline{l}^{\mu} + \beta n_5^{\mu}$. In this case $l \cdot n_6 = 0$.
With this choice, all we need to 
do is to add one more polarization direction to Eq.~(\ref{eq5a5}). We obtain
\be
\sum \limits_{i=1}^{4} e_{i}^{\mu} e_i^{\nu} = 
\left \{ \begin{array}{cc}
-\omega^{\mu \nu} (\overline{l}); & \mu,\nu \in 4{\rm dim};\\
-n_6^{\mu} n_6^{\nu}, & \mu=6, \nu=6, \\
0, & {\rm otherwise}. 
\end{array} \right.
\label{eq6a5}
\ee

We now make the following observation. For a chosen  loop momentum $l$, be it 
four- or five-dimensional, the calculation in $D_s = 6$ differs 
from the calculation in $D_s = 5$ by a single polarization component of 
the gluon, denoted $n_6$ in Eqs.~(\ref{eq6a4},\ref{eq6a5}).
All external 
momenta and polarizations are four-dimensional, and the cut loop 
momentum $l$ that satisfies the unitarity constraint is at most 
five-dimensional. Because of this the  $n_6$ polarization 
gives non-vanishing contribution when it is contracted with itself through 
a metric tensor.
From this point of view, its contribution is equivalent to that of an additional (real) scalar particle in the loop.  To see this more explicitly, 
consider a three-gluon vertex with two gluons  polarized 
along the sixth dimension in six-dimensional space and the third gluon 
with  the four-dimensional polarization 
vector. Since none of the gluon momenta
have a component along the sixth dimension, we obtain
\be
V^{(3)}_{\mu_1 \mu_2 \mu_3}(k_1,k_2,k_3) e_1^{\mu_1} n_6^{\mu_2} n_6^{\mu_3} 
\sim e_1^{\mu_1} (k_2 - k_3)_{\mu_1}.
\ee
The object on the r.h.s. of the above equation is the scalar-scalar-gluon 
vertex. Similar considerations applied to a four-gluon vertex with two gluons 
polarized along the $n_6$ direction immediately lead us to conclude that 
it becomes a scalar-scalar-gluon-gluon vertex. Hence, any tree 
$n$-gluon amplitude with two gluons whose polarization vectors are 
$e^\mu = n_6^\mu$ is equivalent to a tree scattering amplitude 
of $n-2$ gluons and two scalars. This allows us to write
\be
{\cal A}_{(D,D_s= 6)} = 
{\cal A}_{(D,D_s = 5)}
+{\cal A}_{(D)}^{S},
\label{eq12}
\ee
where the amplitude ${\cal A}_{D}^{S}$ is computed with a particular 
propagator for all internal gluon lines $-in_6^\mu n_6^{\nu}/l^2$ which, 
as we argued above, is equivalent to  scalar contribution.

If we use Eq.~(\ref{eq12}) in Eq.~(\ref{eq29}), we derive the following result
\be
{\cal A}^{\rm FDH} = {\cal A}_{(D,D_s=5)} - {\cal A}_{(D)}^{S}.
\label{eq13}
\ee
It tells us that, from the point of view of generalized 
unitarity, computations 
of gluon scattering amplitudes 
in the four-dimensional helicity scheme are equivalent 
to  calculations of gluon scattering amplitudes in 
five-dimensional space-time 
up to a term that can be associated with a scalar contribution.

Finally, we may now write explicitly the relation between amplitudes 
computed in FDH and HV schemes.
Using Eq.~(\ref{eq2.4}) and Eq.~(\ref{eq12}), we derive
\be
{\cal A}^{\rm HV} = {\cal A}_{(D,D_s=5)} - (1+2\ep) {\cal A}_{(D)}^{S}\ .
\ee
We see that the difference between the HV and FDH schemes is entirely due 
to the additional ``scalar'' degree of freedom that contributes to one-loop amplitudes.
Note that we re-derive the results of~\cite{BDKOneloopInt} without resorting
to SUSY arguments. 
The relation between unrenormalized amplitudes in two schemes becomes
\be
{\cal A}^{\rm FDH} -
{\cal A}^{\rm HV}  
= 2\ep {\cal A}^{\rm S}.
\ee
The contribution of a real scalar to one-loop 
gluon scattering amplitudes reads
\be
{\cal A}^{\rm S} = \frac{c_\Gamma}{6\ep} {\cal A}_{\rm tree} + {\rm finite\ }\ ,
\ee
where
\be
c_\Gamma = \frac{\Gamma(1+\ep) \Gamma(1-\ep)^2 \mu^{2\ep}}{(4\pi)^{2-\ep}}\ ,
\label{eq4.1a}
\ee
is the usual normalization factor and 
$\mu$ is  the scale that maintains 
the dimensionality of loop integrals after the loop momentum is continued 
to $D=4-2\ep$ dimensions. 
Neglecting terms of order $\ep$, we obtain the well-known relation 
between gluon amplitudes computed in FDH and HV schemes~\cite{bdk}
\be
{\cal A}^{\rm HV} = {\cal A}^{\rm FDH} - \frac{c_\Gamma}{3} {\cal A}_{\rm tree}\ .
\label{eq125}
\ee

\section{Results}
We now present the results of the calculation of four-, five- and six-gluon 
scattering amplitudes in QCD. If, for a given choice of gluon helicities, 
the corresponding tree amplitude does not vanish, the $n$-gluon 
one-loop sub-amplitude is written in the following way
\be
{\cal A}_{n,1}(1^{\lambda_1},\ldots ,n^{\lambda_n})
= c_\Gamma  \left  ( 
-\frac{n}{\ep^2} 
+ \frac{1}{\ep}\left (-\frac{11}{3} + \sum \limits_{i=1}^{n} \ln (-s_{i,i+1})  \right )
+ \Delta_{\lambda_1,\ldots, \lambda_n} \right) 
{\cal A}_{\rm tree}(1^{\lambda_1},\ldots ,n^{\lambda_n}),
\label{eq4.1}
\ee
where $s_{i,i+1} = 2p_i \cdot p_{i+1} + i\delta$, $p_{n+1} = p_1$. 
The convenience of representing 
scattering amplitudes as in Eq.~(\ref{eq4.1}) is that phase conventions 
for gluon polarization vectors drop out when functions 
$\Delta_{\lambda_1,\ldots ,\lambda_n}$ are computed;
 this feature allows for direct comparison with 
the literature.

We have verified that our calculations
correctly reproduce the divergent parts of 
Eq.~(\ref{eq4.1}). In Tables~\ref{4g},~\ref{5g} and~\ref{6g} we give the results
for finite parts of color-ordered 
sub-amplitudes $\Delta_{\lambda_1,\ldots ,\lambda_n}$. The
finite parts are 
split into  cut-constructible and rational parts 
$$
\Delta_{\lambda_1,\ldots ,\lambda_n} =
\Delta_{\lambda_1,\ldots ,\lambda_n}^{\rm cut} 
+ \Delta_{\lambda_1,\ldots ,\lambda_n}^{\rm rat}  
$$
for kinematic points specified below.  

On the other hand, some  color-ordered 
sub-amplitudes for $n$-gluon scattering are finite; 
the corresponding tree amplitudes vanish making 
representation Eq.~(\ref{eq4.1}) senseless. For those finite amplitudes
explicit results are presented below. Note however that in this case 
results depend on phase conventions adopted for gluon polarization vectors.
To get rid of this dependence,
for finite  amplitudes we compare absolute values $|{\cal A}_{n,1}|$ to 
the results available in the literature. 
All results for scattering  amplitudes reported below are given 
in the FDH scheme; results in the HV scheme can be obtained using Eq.~(\ref{eq125}).

Finally, we note that 
QCD sub-amplitudes are often calculated using supersymmetric 
decomposition since in this way analytic computations are simplified.
In particular, the only part of the gluon scattering 
amplitude that cannot be obtained 
from four-dimensional unitarity is the part where gluons scatter through 
a loop with virtual scalars. Dealing with virtual scalars is   
easier  than with virtual 
gluons since the number of degrees of freedom is smaller.
However, since our goal is to demonstrate the vitality of the method, 
we do not employ this simplification and compute  sub-amplitudes, 
without using the supersymmetric decomposition.

\subsection{Four gluon scattering}

We consider color-ordered 
sub-amplitude ${\cal A}_{4,1}(1^{\lambda_1},\ldots ,4^{\lambda_4})$ 
for the following choice of  external momenta ($p = (E,p_x,p_y,p_z)$)
\ba
&&p_1=\left ( 1,0,0,1 \right),\;\;\;p_2=\left ( 1,0,0,-1 \right), \nonumber \\
&&p_3=\left ( -1,\sin\theta,0,\cos\theta \right),\;\;\
p_4=\left (-1,-\sin\theta,0,-\cos\theta \right ).
\ea
with $\theta=\pi/3$. 

\begin{tiny}
\begin{table}[th]
%\vspace{0.1cm}
\begin{center}
\begin{tabular}{|c|c|c|c|}
\hline\hline
$\lambda_1,\lambda_2,\lambda_3 ,\lambda_4$ & $\Delta^{\rm cut}$ & $\Delta^{\rm rat}$& $\Delta$ \\ \hline\hline
-\;-\;+\;+ & 2.53627 & 0.2222 & 2.75849 \\ \hline
-\;+\;-\;+ &1.90292-3.29626\;i& 0.66667& 2.56959-3.29626\;i \\ \hline \hline
\end{tabular}
\caption{\label{4g} Finite parts of singular four-gluon scattering 
amplitudes for various gluon helicities. 
Cut-constructible and rational parts are shown separately. }
\vspace{-0.1cm}
\end{center}
\end{table}
\end{tiny}

There are four  sub-amplitudes that we have to consider $++++$,$-+++$,
$--++$ and $-+-+$. The amplitudes $++++$ and $-+++$ are finite, i.e. the 
cut-constructible parts of these amplitudes vanish 
identically and the entire results 
are due to rational parts. We find
\ba
&& {\cal A}_{4,1}(1^+,2^+,3^+,4^+) = i\;c_\Gamma \times\;0.33333,
\nonumber \\
&& {\cal A}_{4,1}(1^-,2^+,3^+,4^+) = i\;c_\Gamma \times\;0.7500.
\label{eq4.2}
\ea
Finite parts  for the two divergent amplitudes $--++$ and $-+-+$ are given in 
Table~\ref{4g}. The results for scattering amplitudes in 
Eqs.~(\ref{eq4.1},\ref{eq4.2})
and Table~\ref{4g} are in complete agreement with Ref.~\cite{kst}.

\subsection{Five-gluon scattering}
We consider sub-amplitude ${\cal A}_{5,1}(1^{\lambda_1},\ldots ,5^{\lambda_5})$ 
and choose external momenta to be
\ba
&&p_1=\left ( 1,0,0,1 \right),\;\;\; p_2=\left ( 1,0,0,-1 \right),  \\
&&p_3=\xi \left ( -1,1,0,0 \right ),\;\;\; 
p_4= \xi \left (-\sqrt{2}, 0, 1, 1 \right ), \nonumber \\
&&p_5= -p_1 - p_2 - p_3 - p_4, \nonumber
\ea
with $\xi = 2/(1+\sqrt{2} + \sqrt{3})=0.4823619098$.
We have to consider four  sub-amplitudes $+++++$, $-++++$, $--+++$ and
$-+-++$. The first two amplitudes are finite and, hence, are entirely due to 
rational parts.

\begin{tiny}
\begin{table}[th]
\vspace{0.5cm}
\begin{center}
\begin{tabular}{|c|c|c|c|}
\hline\hline
$\lambda_1,\lambda_2,\ldots,\lambda_5$ & $\Delta^{\rm cut}$ & $\Delta^{\rm rat}$& $\Delta$ \\ \hline\hline
-\;-\;+\;+\;+ & 7.3382230-2.1860355\;i  & 0.24488559-1.4089423\;i& 7.58310859-3.5949778\;i \\ \hline
-\;+\;-\;+\;+ & 12.059206+1.7853279\;i  & -7.5603579-8.4763597\;i & 4.4988481-6.6910318 \;i \\ \hline \hline
\end{tabular}
\caption{\label{5g} Finite parts of singular five-gluon scattering 
amplitudes for various gluon helicities. 
The cut-constructible and rational parts are shown separately. }
\vspace{-0.3cm}
\end{center}
\end{table}
\end{tiny}

For the two finite amplitudes we obtain
\ba
&& {\cal A}_{5,1}(1^+,2^+,3^+,4^+,5^+) = i\;c_\Gamma\times  \left( 0.01056-0.6614\;i \right),
\nonumber \\
&& {\cal A}_{5,1}(1^-,2^+,3^+,4^+,5^+) = i\;c_\Gamma\times  \left( -0.6773-0.4976\;i \right).
\ea
The results for the two divergent amplitudes are given in Table~\ref{5g}.
For all scattering amplitudes, we find complete agreement with the 
results reported in Ref.~\cite{bdk}.

\subsection{Six gluon scattering}

We now present the results for the six-gluon scattering. We consider 
color-ordered sub-amplitude ${\cal A}_{6,1}(1,2,3,4,5)$ for the same external momenta as 
considered in Ref.~\cite{Ellis:2006ss}\footnote{Our conventions differ
from  Ref.~\cite{Ellis:2006ss} in that $x$ and 
$y$ axes are interchanged.}. Specifically, we have
\ba
&& p_1 = 3 \left (-1,\cos \theta \sin \phi,\sin \theta, \cos \theta \cos \phi \right), \;\;\;p_2 = -3 \left (1, \cos \theta \sin \phi,\sin \theta, \cos\theta, \cos \phi \right ); \nonumber \\
&& p_3 =  2 \left ( 1,0,1, 0 \right ),\;\;\;
p_4 = \frac{6}{7} \left ( 1, \sin \beta, \cos \beta,  0 \right ), \nonumber \\
&& p_5 =  \left (1, \cos \alpha \sin \beta, \cos \alpha \cos \beta,  \sin \alpha \right ),\;\;\; p_6 = -p_1 - p_2 - p_3 - p_4 - p_5,
\ea
where $\theta = \pi/4,\phi=\pi/6,\alpha=\pi/3$, $\cos \beta = -7/19$.

\begin{tiny}
\begin{table}[htbp]
%\vspace{0.1cm}
\begin{center}
\begin{tabular}{|c|c|c|c|}
\hline\hline
$\lambda_1,\lambda_2,\ldots ,\lambda_6$ & $\Delta^{\rm cut}$ & $\Delta^{\rm rat}$& $\Delta$ \\ \hline\hline
-\;-\;+\;+\;+\;+ & -19.481065+78.147162\;$i$ & 28.508591-74.507275\;$i$ &9.027526+3.639887\;$i$ \\ \hline
-\;+\;-\;+\;+\;+ &-241.10930+27.176200\;$i$ & 250.27357-25.695269\;$i$ & 9.164272+1.480930\;$i$ \\ \hline 
-\;+\;+\;-\;+\;+ & 5.4801516-12.433657\;$i$ & 0.19703574+0.25452928\;$i$ & 5.677187-12.179127\;$i$ \\ \hline 
-\;-\;-\;\;+\;+\;+& 15.478408-2.7380153\;$i$ & 2.2486654+1.0766607\;$i$ & 17.727073-1.661354\;$i$ \\ \hline
-\;-\;+\;-\;+\;+&-339.15056-328.58047\;$i$ & 348.65907+336.44983\;$i$& 9.508509+7.869351\;$i$ \\ \hline
-\;+\;-\;+\;-\;+ &31.947346+507.44665\;$i$ & -17.430910-510.42171\;$i$ & 
14.516436-2.975062\;$i$ \\ \hline\hline
\end{tabular}
\caption{\label{6g} Finite parts of singular  six-gluon scattering 
amplitudes for various gluon helicities. 
The cut-constructible and rational parts are shown separately. }
\vspace{-0.1cm}
\end{center}
\end{table}
\end{tiny}

We compute  sub-amplitude ${\cal A}_{6,1}(1,2,3,4,5,6)$ for 
the following helicity configurations $++++++$, $-+++++$, $--++++$,
$-+-+++$,$-++-++$, $---+++$, $--+-++$, $-+-+-+-+$.   
Among eight  amplitudes, two amplitudes $++++++$ and $-+++++$
are finite; the corresponding tree amplitudes vanish. 
For these amplitudes we find
\ba
&& {\cal A}_{6,1}(1^+,2^+,3^+,4^+,5^+,6^+) = i\;c_\Gamma\times  \left ( 
0.4968861-0.1838453\;i \right ),
\nonumber \\
&& {\cal A}_{6,1}(1^-,2^+,3^+,4^+,5^+,6^+) =  i\;c_\Gamma\times
\left (
1.0182879+3.0968492\; i \right ).
\ea
The results for the finite parts of remaining six gluon scattering amplitudes 
are given in Table~\ref{6g}, where cut-constructible and rational parts
are shown separately.  For all six-gluon amplitudes, we find 
an agreement\footnote{To obtain numerical results reported  
Ref.~\cite{Ellis:2006ss} from our results, 
one should expand $\mu^{2\ep}$ included  in the normalization factor $c_\Gamma$, Eq.~(\ref{eq4.1a}), in powers of $\ep$ and substitute $\mu = 6$.},
at least through seven digits, 
with the results reported in Ref.~\cite{Ellis:2006ss}.

\section{Conclusions}

In this paper we describe a novel method for calculating 
one-loop scattering amplitudes, including their rational parts. It is 
 based on unitarity cuts in higher-dimensional
space-time. Similar to four-dimensional unitarity,  one-loop amplitudes
are obtained  from tree amplitudes. These tree amplitudes can 
be efficiently calculated
using recursive algorithms of polynomial complexity leading to an
efficient method for semi-numerical evaluation of one-loop amplitudes.
Because the method is built around integer-dimensional unitarity cuts,
we do not foresee any difficulty with its application
to chiral gauge theories.

Our method  solves the outstanding problem of an efficient semi-numerical
evaluation of the so-called rational part of one-loop amplitudes, an important step
 towards automated computation of NLO cross sections.
The generality
of the method allows a straightforward calculation 
of NLO corrections to multi-particle  processes that involve virtual particles
of arbitrary spins and masses.
We hope that further development of
this method will finally bring within reach NLO computations for
such complicated processes as $PP\to t\overline{t} + 2$ jets
and $PP\to V + 3,4$ jets.

\section*{Acknowledgments} 
K.M. is supported in part by the DOE grant DE-FG03-94ER-40833 and Outstanding 
Junior Investigator Award.  We are indebted to G.~Zanderighi for her 
help with comparison of our numerical results to  that of 
Ref.~\cite{Ellis:2006ss}. Useful discussions with R.K.~Ellis are 
gratefully acknowledged.

\end{document}